# PHOTOCONDUCTIVITY AND PHOTOEMISSION OF DIAMOND UNDER FEMTOSECOND VUV IRRADIATION

J. GAUDIN [(1)], G. GEOFFROY [(1)], S. GUIZARD [(1)], V. OLEVANO [(1)], S. ESNOUF [(1)], S.M. KLIMENTOV [(2)], P.A. PIVOVAROV [(2)], S.V. GARNOV [(2)], P. MARTIN [(3)], A. BELSKY [(3)] AND G. PETITE [(1)]

(1) Laboratoire des Solides Irradies, UMR 7642, CEA/DSM, CNRS/SPM et Ecole polytechnique, F-91128 Palaiseau Cedex, France
(2) General Physics Institute of the RAS , 38 Vavilov St., 119191, Moscow, Russia
(3) Laboratoire CELIA, UMR 5107, Universite Bordeaux I, 351 Cours de la Liberation, F-33405 Talence

**Abstract** : In order to gain some insight on the electronic relaxation mechanisms occuring in diamond under high intensity laser excitation and/or VUV excitation, we studied experimentally the pulsed conductivity induced by femtosecond VUV pulses, as well as the energy spectra of the photoelectrons released by the same irradiation. The source of irradiation consists in highly coherent VUV pulses obtained through high order harmonic generation of a high intensity femtosecond pulse at a 1.55 eV photon energy (titanium-doped sapphire laser). Harmonics H9 to H17 have been used for photoconductivity (PC) and harmonics H13 to H27 for photoemission experiments (PES).
As the photon energy is increased, it is expected that the high energy photoelectrons will generate secondary e-h pairs, thus increasing the excitation density and consequently the PC signal. This is not what we observe : the PC signal first increases for H9 to H13, but then saturates and even decreases.
Production of low energy secondary e-h pairs should also be observed in the PES spectrum. In fact we observe very few low energy electrons in the PES spectrum obtained with H13 and H15, despite the sufficient energy of the generated free carriers. At the other end  (H21 and above), a very intense low energy secondary electron peak is observed.
As a help to interprete such data, we realized the first *ab initio* calculations of the electronic lifetime of quasiparticles, in the GW approximation in a number of dielectrics including diamond. We find that the results are quite close to a simple "Fermi-liquid" estimation using the electronic density of diamond. We propose that a quite efficient mechanism could be the excitation of plasmons by high energy electrons, followed by the relaxation of plasmons into individual e-h pairs.

It is well known that diamond is a particularly interesting material in view of various applications ranging from fast-high voltage electronics [1] to VUV detection [2], due to its exceptional carrier mobility, as well as to its excellent thermal and mechanical characteristics. All such applications rely on a good control of the free carriers properties including, for detection of high energy particles, those of hot carriers. A specially interesting problem, far from being solved, is that of the various energy relaxation mechanisms, including (in the case of hot carriers) the electronic relaxation mechanisms. In this paper, we use the opportunity offered by modern ultrafast laser and XUV sources to study this mechanisms using two methods : Transient Photoconductivity (TPC) [3] which, due to its simplicity of use and high sensitivity, allows to study over a wide range of parameters the dependence of the photoinjected free carrier density on various radiation parameters, and particularly in our case the photon energy, and Ultraviolet Photoelectron Spectroscopy (UPS), which offers the possibility of investigating the free carriers energy distribution.
More specifically, we intend in this study to investigate the free carrier multiplication mechanism. If the energy of the photons contained in the pulse used to excite the sample is high enough, the photocarriers it creates are able to excite secondary electron-hole pairs. Needless to say, such processes would have a strong influence on the sensitivity of diamond as a VUV detector, so that they are worth studying not only from the fundamental point of vue, but also because of their impact on thechnological application of diamond.
The question of the threshold energy at which such processes begin to occur is reasonably simple in the case of materials with an idealistic band structure : energy and momentum conservation imposes (for materials with direct band gap parabollic bands) a threshold photon energy $E_p = 2E_G$ for materials with a perfectly flat valence band – trapped holes - and  $E_p = 4E_G$ for a material with symmetric bands. But for diamond, which has an indirect bandgap and several degenerated bands the answer is not so simple. Based on its similarity with Si or Ge, one should expect the threshold photon energy to lie between 3 and 4 $E_G$ [4]. Besides, answering the question of the energy threshold does not bring any information on the efficiency of the process.  Both TPC and UPS should bring quite useful information concerning such questions, provided short pulses of VUV radiation in the relevant energy  range can be used. In the case of diamond, with its direct bandgap close to 7eV, this means photon energies between 21 and 28 eV. It turns out that high order harmonic generation (HHG) of titanium-doped sapphire (ti:sa) lasers (1.55 eV photon energy) precisely produces high intensity femtosecond pulses in this range (one is in the range of harmonics 13 to 19).

*Samples* : The samples used in the experiment are either polycrystalline CVD or IIa single crystals (in the case of TPC, only the latter one having been investigated using UPS), with dimensions 0.4x4x4 mm$^3$. The sample purities have been evaluated using X-band EPR spectroscopy before further investigations. The CVD sample exhibited a quite visible P1 line (nitrogen donnor center : typically a few $10^{16}$ spins.cm$^{-3}$) and a broader line usually attributed in the litterature to "dangling bonds". The IIa sample did not show any signal using X-band, and a very small signal (precluding any serious interpretation, though the g-factor was consistent with a P1 signal, with a density below $10^{15}$cm$^{-3}$) using Q-band spectroscopy. We deduce from this evaluation that both samples, and particularly the IIa one, are of high purity. To eliminate surface conductivity of diamond, all the samples were annealed in oxidizing atmosphere before TPC measurements

*XUV sources* : Two different HHG-XUV sources were used. TPC measurements were performed using the LUCA setup in CEA-Saclay. Harmonics 9 to 19 of an intense ti:sa femtosecond laser were generated in a high pressure Ar or Xe gaz jet. Up to 20 mJ in a 50 fs pulse were focused using a 5m focal length lens, yielding HHG pulses with typically $10^{10}$ photons per pulse (at H15). Harmonic selection was done using a combination of grazing incidence grating and toroidal mirror. The initial harmonic pulse duration (<50 fs) is increased in this case, but was calculated to stay below 1 ps. The laser repetition rate was in this case of 25 Hz, but the TPC measurements are made on a shot by shot basis. UPS measurements were performed using a similar system, but based on a high repetition rate (1kHz) ti:sa system at CELIA laboratory (Univesité Bordeaux I). In this case, harmonics 13 to 27 were used.

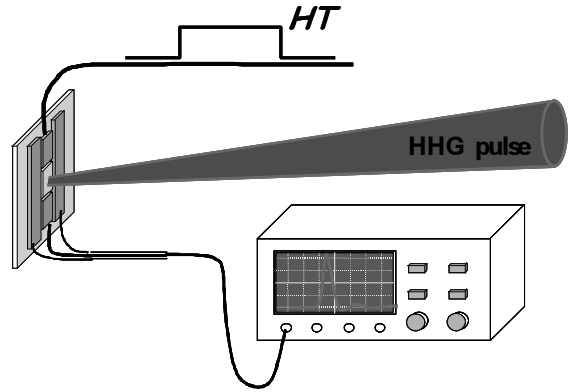

**Figure 1** : principle of the TPC measurements

*TPC measurements* : The TPC technique is based on the measurement of a transient current through diamond samples either inserted into PC cell or located just in front of it, as shown on Fig. 1, and exposed to laser pulses. Detailed description of this method can be found in [3]. We briefly reproduce here some basics to provide better understanding of the results and peculiarities introduced by absorption in diamond in such a wide spectral range. Essentially, only a small area within the sample plates was irradiated (from 30 μm to 2 mm in diameter). The cell consisted of two electrodes without contacts to the sample which, contrary to conventional PC using ohmic connections to the photoconducting material, allowed to measure short peaks of displacement current through the load resistor when a high voltage pulse (up to 3 kV, 50 μs) was applied synchronously with the incident laser pulse. The short TPC trace was caught and stored by a broad-band digital oscilloscope (2 GHz bandwidth). The typical TPC pulse measured in diamond is shown on Fig. 1. Here, information on photo-excitation and conduction processes can be derived from the following parameters: the peak amplitude, its intensity dependence, shape and duration of the rear front. The amplitude voltage of the conductivity pulse is given by an equation

$$U_{pc} = \frac{eU_0 R}{L^2 \varepsilon}(\mu_e N_e + \mu_h N_h)\alpha\varphi(\tau_c/\tau_p) \qquad (1)$$

where $e$, $\varepsilon$, $\mu_e$ and $\mu_h$ are: charge of electron, permittivity of diamond (5.7), drift mobility of generated electrons and holes (1000-2000 cm$^2$V$^{-1}$s$^{-1}$ in the purest samples), the total number of which is given by $N_e$ and $N_h$. Parameters $U_0$, $R$ and $L$ are, correspondingly, the applied voltage (typically 1.5 kV), load resistance and distance between the electrodes (3÷10 mm). The coefficients $\alpha$ and $\varphi$ take into account deviation of the applied field configuration from the homogeneous one ($\alpha$ = 1 for the flat capacitor) and the ability of the measurement circuit, and of the oscilloscope, to transmit and measure TPC pulses as short as $\tau_p$ ($\varphi$ = 1 when the shortest pulses $\tau_c$ acquired by the measurement circuit without distortion of amplitude are shorter than $\tau_p$).

As seen, the amplitude is directly proportional to the number of free carriers induced in the material by laser radiation. Duration and shape of the rear front of the TPC pulse allows to quantify free charge relaxation dynamics, for example, to measure lifetime $\tau_r$ of free electrons and holes in the conduction band. The main limitation to transient PC measurements is imposed by space charge formation in the laser irradiated domain of the sample. This inhomogeneous charge density profile, caused by displacement of free charge carriers and their subsequent trapping, introduces an inner screening field, the direction of which is opposite to the applied external electric field. Obviously, the screening field has to disappear by the time of the next laser shot not do reduce the measured TPC values. Such relaxation of the screening field fortunately happens by itself in diamond, at room temperature and illumination conditions, and takes from few to several tens of seconds, depending on initial concentration of free electrons and holes. The effect of screening field can be reduced by exposing lager spots on the sample surface.

In this experiment, we proceeded by recording a series of $U_{PC}(I_H)$ characteristics for each harmonic between 9 and 19. The intensity $I_H$ of the harmonic was measured with a photomultiplier illuminated by the harmonic beam reflected on the sample. To extract from this series of characteristics a spectral behavior (at a given $I_H$), it is necessary to know both the photomultiplier spectral sensitivity (given by the manufacturer) and the spectral dependence of the sample reflectivity. This can be found in the litterature [5, 6] but with differences depending on the authors so that we also realised our own measurement (for H13 to H17). The spectral behavior we obtained both on CVD and IIa samples are shown on figure 2

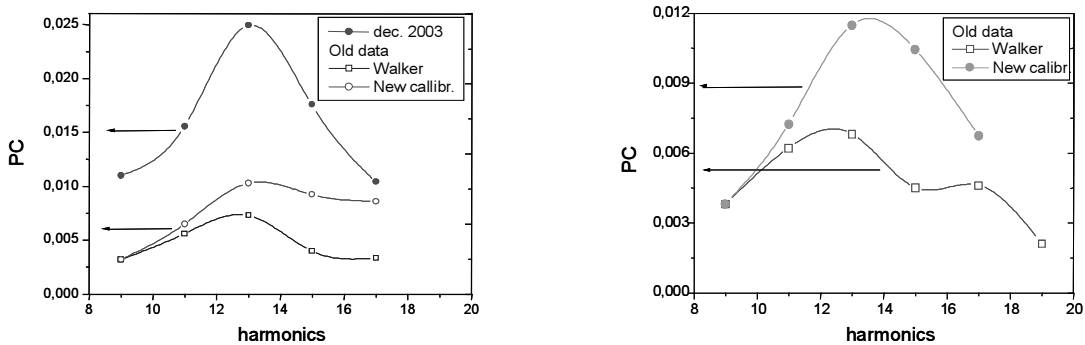

Figure 2 : spectral dependence of the TPC signal for a CVD (left) and a IIa (right) sample.

In the case of the CVD sample, two independent measurements were performed at a several months interval, and one see obvious differences in the sensitivity of the two experiments, as well as differences depending on which calibration we chose for the reflectivity, but a common general behavior can clearly be observed : the TPC signal first strongly increases from H9 to H13, and then diminishes from harmonics 13 to 17 (19 in the IIa case). This is not at all what one would expect on the basis of the "carrier multiplication" argument given above : depending wether one considers the direct or indirect bandgap (a detailed analysis of the momentum consevation should be done before deciding so), "harmonic 14" (but even harmonics are not generated in gasses) would be where the photon energy would cross the 2 to 3 $E_G$ (direct) or 3 to 4 $E_G$ boundary. Concerning the H19 point on the IIa sample, the photon energy is there more than four times the direct bandgap and five times the indirect bandgap. Clearly the simple "energetic" argument (particularly the "flat valence band" model) is not precise enough here to predict the behavior of this specific material. Considering the actual diamond valence band (as obtained from the calculation of [7]) and the possible absorption mechanisms, allows to understand why it may be so (Figure 3)

It should first be underlined here that in TPC experiments with XUV pulses, all the energy entering in the sample is absorbed (each photon giving initially one e-h pair) so that it is essentially the possibility of carrier multiplication that should make the difference. Now, the diamond conduction band is constituted of two groups of bands (we call them CB1 and CB2). The lowest one, CB1, has a total width of about 14 eV, twice the direct bandgap, in principle enough energy to allow secondary excitations provided momentum conservation can be realised. But a first remark is that it is not possible with either H13 or H15 to excite the whole range of CB1, at least using direct transitions. One has to find selected points in the Brillouin zone where resonant conditions can be realised. Taking this into account, we showed on fig. 3 the maximum energy that can be obtained in CB1 by absorption of a H13 (dotted arrow) of H15 (full arrow) photon. One can see that (1) this energy is significantly lower than the total width of CB1 and (2) that it is not very different depending on the harmonic used. Of course, it is possible with H15 to excite also states in CB2, but trying to do so, one realises that only a quite narrow band of states in CB2 can be reached. It should be pointed out also that another band structure calculation of diamond [8] predicts a 5 eV bandgap between CB1 and CB2, in which case CB2 cannot be reached with H15. The net outcome of this exercise is that, despite the quite different photon energy, the actual excitations that can be realised with H13 and H15 do not differ so much.
Note that there are other possibilities that could explain an unexpected behavior, and in particular the fact that the penetration depth of the radiation strongly varies from H9 to H19 (typically from 1.5 to 8 nm) which may have consequences if material properties vary close to the surface, but notwithstanding, it is clear that a detailed analysis of the actual excitation mechanisms is a prerequisite to any interpretation of spectral variations such as the ones shown on fig.2.
*Electronic lifetime calculations* : in particular, it would be extremely useful to dispose of calculations of the spectral density of excited electrons obtained with each harmonic. The recent progresses of the theory of electronic excitations in solids (using Time Dependent DFT – TDDFT - or Many Body Quantum Theory – MBQT) [9] make such a calculation perfectly possible now, but results are not available yet. However, we performed the first *ab initio* calculations of electronic lifetimes in different dielectrics including diamond. In practice, one starts from the determination of the quasi-particle (the correct designation of the "conduction electrons" that we mentioned above) self-energy operator expressed in the "GW" approximation [10]

$$\Sigma^{GW}(r_1, r_2, \omega) = \frac{i}{2\pi} \int d\omega' G^{DFT-LDA}(r_1, r_2, \omega - \omega') W^{RPA}(r_1, r_2, \omega') \qquad (2)$$

where $G$ is the quasi-particle Green's function and $W$ a screened Coulomb interaction (in eq. (2) and hereafter, the upper indices express the approximation in which the quantities are evaluated - 19 special k-points in the irreducible Brillouin zone are used for the evaluation of these quantities, which must include high symmetry points). The self-energy is then evaluated in a number of discrete points on the imaginary frequency axis (12 points on a linear mesh between 0 and 200 eV). An analytical continuation of $\Sigma^{GW}$ into the whole complex plane is then obtained using a Pade approximant $P(\omega)$, since one is interested in quasi-energies that lie close to the real axis. The quasi-particle energy is then obtained by solving the equation

$$E_{QP} = E_{KS} - E_{xc}^{LDA} + P(\omega = E_{QP}) \qquad (3)$$

where $E_{KS}$ and $E_{xc}$ are respectively the Kohn-Sham and exchange-correlation energies. Finally, the quasi-particle lifetime is calculated as

$$\tau^{-1} = 2\Im(E_{QP}) \qquad (4)$$

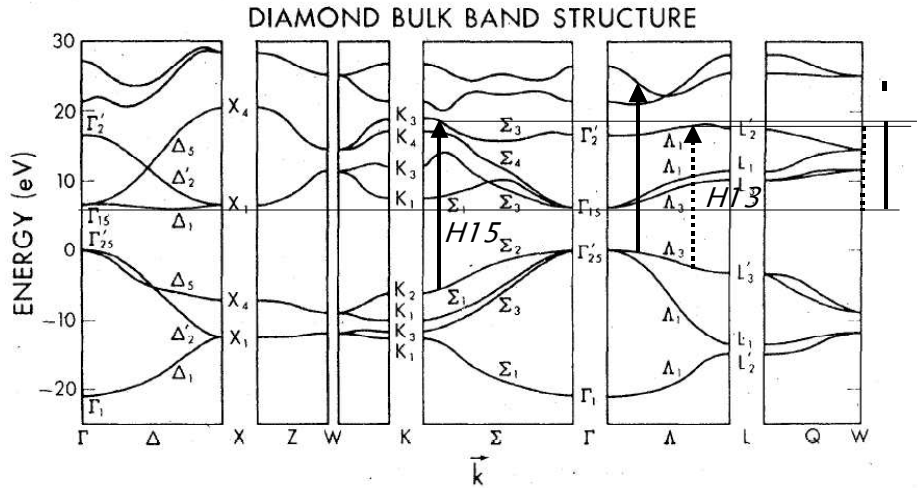

Figure 3 : diamond bulk band stucture (as taken from [7]) and excitation paths using H13 and H15

Since we are neglecting the ionic degrees of freedom in the self-energy (2), the resulting lifetime accounts only for electronic decay channels (e.g. decay by emission of plasmons, e-h pairs, Auger processes). Phonon emission decay and scattering with defects and impurities are neglected. Therefore the lifetime is meaningful only where electronic processes dominate, i.e. far away from the Fermi level. Figure 4 shows the results of such a calculation in the case of diamond, where we concentrated on the 10-50 eV region, which is of interest here (the origin of energies is taken at the top of the valence band), because electronic scattering dominates. The results are compared with the predictions of a Fermi liquid model using the parameters of diamond (and where we have used for zero of energy the bottom of the CB for the electrons, and the top of the VB for the holes), with which they show significant differences, but staying within the same order of magnitude. The purpose of this figure is to make two remarks :

- the electronic lifetime is very short, typically 0.5 fs at a 10 eV of energy in the conduction band (bandgap subtracted from GW energies of fig. 4). This allows a rough estimation of the electron mean free path (using free-electron parameters) which is of the order of the nm. Using the more correct parameters (which requires to enter into the details of the diamond band structure) one can assume that this quantity will somewhat increase (the electron effective mass is smaller than unity) but by less than one order of magnitude. In the following experiment, we will use harmonics whose penetration depth ranges from typically 5 to 15 nm. We can then safely assume that many of the electrons composing the photoemission spectrum (but clearly not all of them) will have been subject to at least one interaction. One thus expect the electron spectra to be a combination of the direct excitation spectrum and of relaxed electrons.
- the general behaviour is not very different from the Fermi liquid one, but some differences are clearly visible. We can identify four regions with a rather regular behaviour, with three transition zones : at 20 eV, a sudden increase (typically 20%) of the lifetime, and a somewhat smaller at 26 eV, which we attribute to band structure effects. More important for us here is the rather rapid decrease (by about 30%) observed around 35 eV, which, once the bandgap energy retrieved, corresponds almost exactly to the diamond Drude plasmon energy (31.5 eV). In fact this transition region exactly coincides with the plasmon peak found in an RPA calculation of the EELS spectrum, but one should remember that such a calculation is made for a high electron initial energy, so that it does not directly apply to the situation we investigate. But the lifetime calculation does, and it also shows an quite significant effect in this region.

Let us finally note that alternate integration methods can be used. They give results that differ significantly from the above results in the low energy region (where the "Fermi Liquid" behavior becomes quite questionable), but the essential features discussed above subsist.

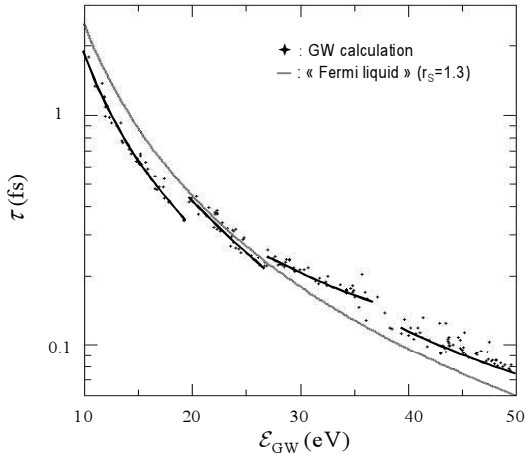
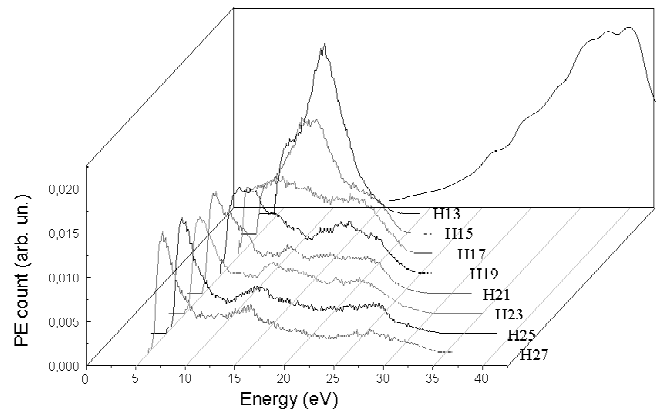

Figure 4 : GW electronic lifetime of excited electrons in diamond. The continuous curve shows for comparison the prediction of a "Fermi liquid" type calculation, using the electronic density of diamond. The black lines are added by hand as a guide to the eye.

Figure 5 : UPS spectra of diamond obtained with harmonics 13 to 27 of a Ti/saph femtosecond laser. The spectrum on the back panel is the result of the EELS calculation mentioned in the text. For each spectrum the line is extended up to the position of the maximum energy expected (photon energy).

*Femtosecond UPS experiments* : we performed UPS measurements of the electron energy distribution (as explained above, at least partially relaxed) using the harmonics H13 to H27 of the CELIA laser and a 150 mm radius hemispherical analyser (resolution better than 0.1 eV). The setup is a traditional one for such experiments on synchrotron radiation sources, the only specific points being the VUV source, and some specific precautions that need to be taken in the case of insulating samples (essentially here working at a high – 400 C – sample temperature : for further details see [11]). The spectra obtained are presented on figure 5. Let us mention that in our experimental conditions, there are some uncertainties concerning the contact potentials on one side, and also that we cannot expect our surface to exhibit the negative affinity that was measured on (100) diamond surfaces after annealing to very high temperatures [12]. The spectra have been aligned in order to set their onset at the position of the theoretical indirect bandgap (the origin of energies being taken at the top of the valence band). Therefore the uncertainty on the exact position of the spectral features could easily reach the eV. We thus will concentrate below on features that do not require such a precision. Concerning the intensity, the spectra have been normalized to a constant total photoemission signal. One can make the following remarks :
- for low harmonics (H13 and H15) one does not observe a significant low energy peak while such a peak is clearly obsevred for the higher harmonics (H19 and above). Moreover the low harmonic spectra, apart from the absence of the "negative affinity peak" agree rather well with those measured in [12].

From the results reported there, one could estimate that our spectra should be shifted up by aprrox. 1 eV, which has basically no consequence on what follows.
- for high harmonics, there is a significant electron deficit on the high energy side compared to the expectations : for harmonics 23 and above there is a region as broad as 5 eV where no electrons are detected even though the photon energy allows excitations in this region.
- the width of the valence band is 20 eV, so the width of the excited region in the conduction band should no exceed this value. As a result all the electrons detected for, e.g., H27 between 5 and 22 eV are necessarily the result of a relaxation process. They either result of secondary excitations or from the relaxation of higher energy electrons.
- both the appearance of the low energy peak and the disappearance of the high energy electrons show up when the maximum attainable energy enters in the region of the plasmon loss.
These results lead us essentially to the two following conclusions (1) in the case of low harmonics, we do not find here again a clear evidence of the electron carrier multiplication mechanism (2) the appearance of a strong relaxation is correlated with the possibility of exciting plasmons. As a result, collective mechanisms such as the excitation of a plasmon which further desintegrates into individual pair excitations could be a quite efficient relaxation mechanism (which is known to occur in metals).

In conclusion, we have study using two very different methods the dynamics of photoinjection of garriers by ultrashort XUV photons in diamond. Our results, when compared to simple models concerning the question of carrier multiplication through secondary e-h pair excitations show a quite different behavior. Clearly the subtelties of the material band structure should be taken into account more precisely to make predictions on this point. A theoretical study of the electronic lifetimes showed that surprisingly enough, the Fermi liquid model (that is not supposed to be applicable to insulators) gives very reasonable orders of magnitude and tendencies, even though some features related with band structure effects and plasmon excitations can be identified. One also realizes that the interpretation of UPS spectra in this VUV range is not a simple matter, and that we could learn a lot from the new developments of *ab-initio* calculations which could at least provide reliable photoexcitation spectra.

The autohrs acknowledge the help of grants INTAS 01-458, BFBR-03-02-16725 and PAI-04527 ZM. JG acknowledge the help of Region Aquitaine.